\documentstyle[pre,aps,preprint]{revtex}
\begin{document}
\draft
\title{Random Walk Approach to Simple Evolution Model }
\author{\it L.~Anton\\
Instituto Nazionale di Fisica della Materia,\\
\it International School for Advanced Studies\cite{a} (SISSA),\\
 \it via Beirut, no. 2-4, 34013 Trieste, Italy\\
\it and\\
\it Institute of Atomic Physics, IFTAR, Lab. 22,\\
 \it P.O.Box MG-7, R-76900, Bucharest, Romania}
\maketitle
\begin{abstract}
The dynamics of the avalanche width in the evolution model 
is described using a random
walk picture. In this approach  
the critical exponents  for avalanche distribution, $\tau$,
and avalanche average time, $\gamma$, are found to be the same as in
the previous mean field approximation but SOC appear at
$\lambda_{critical}=2/3$, which is very close to numerical  
value.
A continuous time random walk is studied numerically as a possible way
to  reconstruct in simpler concepts the evolution model.
\end{abstract}
\section*{Introduction}
Self Organized Criticality (SOC) theory  studies non-equilibrium 
spatially extended 
systems  which show scale invariance on a large spatial or temporal 
domain.  The phenomena
which  have these characteristics appear in 
astrophysics, geophysics, biological
 evolution, stock market, etc. ~\cite{B},~\cite{BS},~\cite{pmb} .  
The study of this type of problems has emerged through various 
 models;  we will concentrate to the Bak \& Sneppen
model (BS) \cite{BS} proposed as explanatory for certain aspects of 
the
 biological evolution.
  Its mathematical simplicity attracted numerical studies
\cite{BS},~\cite{Stanley},~\cite{Grass} and mean-field analytic
treatments \cite{SB},~\cite{Derrida}.

The model treats a number of $N$ species interacting on an one 
dimensional
chain \mbox{(a simple pictures of the food chain)}.
  Each species has   assigned a  scalar parameter, called fitness, 
with values in $(0,1)$ interval as a measure for the 
adaptability of species to the ecosystem.
 The dynamics starts from the site with the smallest fitness:
 new random values from $(0,1)$ interval  are independently
attributed to this site and to its two neighbors with a uniform 
distribution and uncorrelated from previous values then the cycle
is repeated for a large number of times. 
For this system  we  define a $\lambda $-avalanche ( $0<\lambda < 1$)
 as the number of steps between two consecutive configuration with 
all the 
fitness values greater than $\lambda$.
Numerically it has been found that the the stationary state exhibits 
scaling
 laws in $N\rightarrow \infty$ limit 
for the correlation function
 of the active sites and avalanches distribution at $\lambda\approx 
 0.66$
 \cite{BS}.
 We mention that the model shows also critical behavior in the 
versions with
more than one dimensions or with slightly modified dynamics
\cite{Stanley}, \cite{Grass}. 
 
In this paper we propose a new approximated solution. Our treatment 
brings in
 the dynamics of the  avalanche width in a mean field approach.
In the BS model the the avalanche width is a random variable which
shows memory effect; we removed this effect by updating all the sites
within an avalanche (see section II for more details), in this way we 
we kept track of the spatial extension of the avalanches, thing which
is not possible  in the infinite range approximation
\cite{Derrida}. Even the
approximation seems too crude we found the value of
$\lambda_{critical}=2/3$, which is very close to the numerical result.

The article organized as follow:
Section I presents the general master equation for the fitness 
distribution and 
the derivation of the mean field equation found with 
probabilistic arguments
in Ref. \cite{BS}.
Section II introduces  our analytical treatment  based on a 
mapping in a random
 walk problem.   In this approximation
we compute exactly the value of $\lambda_{critical}$ and the 
critical exponents
 $\tau$ and $\gamma$ as they were defined in Ref. \cite{Stanley}.
Section  III presents a way to improve the method introducing
a continuous time random walk description, numerically we show 
the improvement
of the critical exponent $\tau$ keeping $\lambda_{critical}$ fixed.
The details of calculation are given in the Appendix.

\section{The Master Equation.}

The BS model is completely characterized by the  probability 
$P(x_1,x_2,\dots ,x_N;t)$
 to find the system in the state $(x_1,x_2,\dots,x_N)$ at the time
 $t$
 given the initial distribution at $t=0$. Because there are not 
memory effects,
 the evolution of the system is described 
by the following master equation
\begin{eqnarray}\label{markov}
P(x_1,x_2,\dots,x_N;t+1) =  \sum_i \int dx_i'dx_{i-1}'dx_{i+1}'
P_{st}(i;x_1,\dots,x_{i-1}',x_i',x_{i+1}'\dots,x_N)\nonumber \\ 
\times P(x_1,\dots,x_{i-1}',x_i',x_{i+1}'\dots,x_N;t),
\end{eqnarray}
where periodic boundary conditions were assumed
and  $P_{st}(i;x_1,\dots,x_{i-1}',x_i',x_{i+1}'\dots,x_N)$ is the 
probability to have activity at site $i$ if the system is in
the  configuration 
$(x_1\dots,x_{i-1}',x_i',x_{i+1}'\dots x_N)$. 
For the original one dimensional BS model 
\begin{equation}
P_{st}(i;x_1,x_2,\dots,x_N)=\prod_{j\neq i}\theta(x_j-x_i),
\end{equation}
where $\theta(x)$ is the step function
\begin{displaymath}
\theta(x)=\cases{1,&if $x> 0$;\cr
                   0,&if $x\le 0$.\cr}
\end{displaymath}
At stationarity, integrating in $(1)$ over $x_2,\dots,x_N$ we get
 easily
 the following relation:
\begin{equation}\label{consv}
P_{ac}(1,x_1)+P_{ac}(2,x_1)+P_{ac}(N,x_1)-{3\over N}=0;
\end{equation}
where
\begin{eqnarray}
P_{ac}(i;x_j)=\int dx_1\dots dx_{j-1}dx_{j+1}\dots dx_N 
P_{st}(i;x_1,\dots, x_N)
\nonumber\\ 
\times P(x_1,\dots,x_N)\nonumber
\end{eqnarray}
 is the the probability to have activity in site $i$ when in site $j$ 
the 
fitness  has the value $x_j$.
If in equation (\ref{consv}) we try a stationary self-consistent 
mean field solution of the form $p(x_1,\dots,x_N)=p(x_1)p(x_2)\dots
 p(x_N)$
 after some algebra we get
\begin{equation}\label{bak}
\biggl(1-{2\over N-1}\biggr)Q^N(x)+{2N\over N-1}Q(x) +3x-3=0
\end{equation}
with $Q(x)=\int_{x}^1p(x')dx'$.
 Equation (\ref{bak})  was previously obtained  in 
Ref.\cite{SB}, in the $N\rightarrow\infty$ limit
one finds $p(x)=3/2\quad x\in(\lambda_{critical},1)$ and $p(x)=0 
\ \mbox{when}
\quad x\in (0,\lambda_{critical}), \ \lambda_{critical}=1/3$  
 whereas  numerically  $\lambda_{critical}\approx 2/3$ \cite{BS}.
The statistical independence between the sites in the 
mean field solution allows the reduction of the problem
 at an one dimensional random walk on the positive semi-axis 
where the state $n$ represents the state of the system 
with $n$ fitness values greater than $\lambda$. The solution 
developed in Ref.
\cite{Derrida} gives the same $\lambda_{critical}$ as predicted by 
eq.(\ref{bak}) and the critical exponents $\tau=3/2,\ \gamma=1$.
\section{The New   Approach.}
We remember here that the 
size of an $\lambda$ avalanche is the number of steps between 
two consecutive
events  with no fitness below the the value $\lambda$, so, it is
a quantity characterizing time intervals. 
For a system of size $N$, with free boundary conditions,
we define the avalanche width  at a given moment $t$ as the 
number of
sites between the most left species with the fitness less than
 $\lambda$ and the most right  species with the fitness less than
 $\lambda$,
the species between these two sites can have any 
value of the fitness. This is a quantity which characterize the
spatial structure of our system.
The width  shows memory effects 
for the originally proposed dynamics,
in the spirit of the mean-field approximation we approximate the
evolution of the avalanche width with the following dynamics: 
 at every step the species between
 right and left extrema are performing 
uncorrelated movements and  we introduce in dynamics the right 
nearest neighbor
of the right extremum species in order to mimic the evolution.
The complete randomness makes the movements of the two extrema
completely equivalent and for this reason we have chosen to move
only in one direction.
In origin we also accept the double
step.  

 With this change the avalanche width  is a random variable 
without memory effects on a discreet
set of states which now can be extended to the entire non-negative
 semi-axis with the state zero corresponding to the state
 with no species below $\lambda$ and the state $n$ to a realization
 of BS
model with with $n$ sites between the most left and the most 
right sites with
 the value of their corresponding fitness less than $\lambda$.
 The transition matrix of the model has the following form:
 \begin{equation}\label{p}
p=\pmatrix{(1-\lambda)^2&2\lambda (1-\lambda)& \lambda ^2&0&0&\ldots
\cr
    (1-\lambda )^2 &2\lambda (1-\lambda)&\lambda ^2&0&0&\ldots\cr
(1-\lambda )^3 &3\lambda (1-\lambda)^2&2\lambda^2 (1-\lambda)&
\lambda ^2&0&
\ldots\cr
(1-\lambda )^4 &4\lambda (1-\lambda)^3&3\lambda ^2(1-\lambda)
^2&2\lambda^2 (1-\lambda)&\lambda ^2&\ldots\cr
\vdots&\vdots&\vdots&\vdots&\vdots\cr}.
\end{equation}
The formulae for the matrix elements are:
\begin{eqnarray}
p_{00}&=&(1-\lambda)^{2},\ p_{01}=2\lambda(1-\lambda),\ p_{02}=
\lambda^2,\
p_{0j}=0,\ j>2;\nonumber\\
p_{j0}&=&(1-\lambda)^{j+1},\quad j\ge 1;\nonumber\\
p_{j1}&=&(j+1)\lambda(1-\lambda)^j,\quad j\ge 1;\nonumber\\
p_{j2}&=&j\lambda^2(1-\lambda)^{j-1},\quad j\ge 1;\nonumber\\
p_{jl}&=&\left\{ \begin{array}{ll}
               p_{j+1,i-1} & \mbox{if $l\ge 2$ and $j\le l-1$};\\
               0 & \mbox{otherwise}
\end{array}
\right.
\end{eqnarray}
with the convention that $p_{ij}$ is the transition probability from 
state
 the $i$ to the state $j$ and  $i,j\in \{0,1,2,\dots\}$.
The distribution probability of avalanches is the first return
 probability
 distribution for this random walk and it  can be written  as
\begin{equation}\label{avp}
p(n+2)=\sum _{i=1}^{\infty} p_{01} \tilde p_{1i}^{(n)}p_{i0}+
\sum_{i=1}^{\infty}p_{02}\tilde p_{2i}^{(n)}p_{i0}
\end{equation}
where $\tilde p_{ij}^{(n)}$ is the $i,j$ element of the $n$-th power 
of the matrix $\tilde p$ obtained
from the matrix $p$ removing the row and the column zero and it is
 describing the
 evolution of the random walk outside of the origin.
The first (second) term in the r.h.s. of eq.(\ref{avp}) represents
 the first
return probability, after $n$ steps, 
when the initial step is single, (double).
For a site different from the origin the forward step can only be 
 single, as matrix $p$ shows.

We modify the first two columns of the transition matrix $p$ such
 that to have 
 the same elements on the diagonals of the $\tilde p$ matrix.
 Keeping the closure relation $\sum_j \tilde p_{ij}=1$ we 
produce the following matrix:
\begin{equation}\label{p'}
p'=\pmatrix{(1-\lambda)^2&2\lambda (1-\lambda)& \lambda ^2&0&0&
\ldots\cr
    (1-\lambda )^2(1+2\lambda) &2\lambda ^2 (1-\lambda)&\lambda ^2&
0&0&\ldots\cr
(1-\lambda )^3(1+3\lambda) &3\lambda^2 (1-\lambda)^2&2\lambda^2 
(1-\lambda)
&\lambda^2&0&\ldots\cr
(1-\lambda )^4(1+4\lambda) &4\lambda^2 (1-\lambda)^3&3\lambda
 ^2(1-\lambda)^2&2\lambda^2 (1-\lambda)&\lambda ^2&\dots\cr\vdots&
\vdots&
\vdots&\vdots&\vdots&\ddots
\cr}
 \end{equation}
The asymptotic behavior of the first return time distribution is the 
same for the
 both random walks described by the matrices $p$ and $p'$.
 In fact in eq. (\ref{avp}) we can go a step further developing 
$\tilde p_{1i}^{(n)}$ with respect to site $1$  the remaining 
matrix from the 
second column and second row is identical with the $\tilde p'$
 matrix obtained
from $p'$ in the same way as $\tilde p$ from $p$. 
\begin{equation}\label{equiv}
p_{1i}^{(n)}=\sum_{n_1+\dots n_j\atop =n-n'}\prod_{l=1}^{j}
\biggl(\sum_{k=2}^{\infty} 
\tilde p_{12}
{\tilde p}^{(n_l)}_{2k}\tilde p_{k1}\biggr)\biggl((1-\delta_{1i})
 \tilde p_{12}
\tilde p_{2i}^{(n')}
+\delta_{1i}\tilde p_{11}^{(n')}\biggr)
\end{equation}
with  $n=n'+n_1+\dots n_j$ and $\delta_{ij}$ the 
Kronecker symbol.  
The terms in eq. (\ref{equiv})  represent multiple returns in 
the site 
$1$ before the last step to the origin.
In the $n\rightarrow\infty$ limit the terms with $n'$ and all 
the $n_j$ bounded but one 
have the same asymptotic behavior as 
${\tilde {p'}}_{1i}^{(n)}$ because they are generated by the
 same matrix;
the other terms will decay exponentially, due to 
 $\tilde p_{12}^{j}$ factor, or as a  power of the leading 
term when there are two ore more unbounded  exponents. 
If $i=1\ n'$ has to be bounded to avoid the exponential decay.
 
In the Appendix we present the computation for  
the generating function of the avalanches distribution probability
(\ref{R}). In terms of the generating function $R(z)=\sum_{t}
z^tP(t)$
 the average time for the avalanche distribution is
\begin {equation}
\bar t={dR(\xi)\over dz}\bigg|_{z=1}={dR(\xi)\over d\xi}{d\xi \over dz}
\bigg|_{z=1}.
\end{equation}
>From  (\ref{R}) we obtain  that  the average time of an avalanche can
be written as
\begin{equation}\label{tau}
\bar t\approx \biggl|\lambda-\lambda_{critical}\biggr|^{-\gamma}
\end{equation}
when $|\lambda-\lambda_{critical}|\ll 1$,
with the  critical exponent $\gamma=1$ 
and the  critical value of $\lambda,\ \lambda_{critical}=2/3$. 
We can also compute the asymptotic behavior of the avalanches 
probability distribution from  the general formula
\begin{equation}\label{asym}
p(t)={1\over (t+1)!}{d^{t}R(\xi(z))\over dz^{t}}\bigg|_{z=0}
=\frac{1}{2\pi i}\oint_{\Gamma}dz{R(\xi(z))\over (z-z_{0})^{t+1}}
\bigg|_{z_{0}=0}
\end{equation}
where $\Gamma$ is an integration contour  in the complex plane
around $z_0$ which do not circle other poles.
 For $\lambda=2/3$ we found
\begin{equation}
 p(t)\approx t^{-\tau},\quad t\rightarrow \infty, 
\end{equation}
with $\tau=3/2$, and exponential decay for  $\lambda \ne 2/3$.
In the previous equations  the critical exponent  $\tau=3/2$
and $\gamma$
have values as obtained in the mean field solution \cite{Derrida},
whereas
$\lambda_{critical}=2/3$ is in extremely good agreement with
the critical
value of $\lambda$ found in numerical experiments \cite{BS},
\cite{pmb}, \cite{Grass}.

 In language of Markov chain one  can say that $\lambda=2/3$
  is the transition point
  between persistent states ($\lambda\leq 2/3$) and transient states 
($\lambda > 2/3$) \cite{Feller1}.
Nevertheless $\lambda$ is not a dynamical parameter for BS model,
 it introduces
 an "observational window"
for a certain variable which we may choose  from the set of
statistical variables   
compatible with the dynamics of the BS model. SOC appears when
there is at least 
one statistical variable with events at all scale lengths.
In our approach $\lambda $-avalanches
are bounded to origin for $\lambda < 2/3$ and they escape 
to $\infty$ for $\lambda >2/3$. At $\lambda = 2/3$ we have the
peculiar  stationary state in which the average time of 
avalanches is diverging,
therefore there are events on the all time scales. 
\section{Improving mean field}
A significant difference between random walk proposed in the
previous section and the BS model consists 
in the fact that in the latter the system 
will spend a characteristic number of  steps in a given  
state because the activity 
can appear   between the most left  and the most right
 sites where the  fitness is less than $\lambda$.
One possible way to improve our approximation is to promote the
 previous random walk to a continuous time random 
 walk with inhomogeneous waiting time distributions,  each waiting
 time 
distribution allowing for the persistence of a given size avalanche.
The general equation for such a process can be written \cite{Feller2} 
\begin{equation}
P_{ik}(t)=\delta_{ik}e^{-c_i t}+\sum_{j=0}^\infty\int_0^tc_ie^{-c_it}
p_{ij} P_{jk}(t-t')dt'
\end{equation}
where $P_{ik}(t)$ is the probability density to have the 
walker in state $k$ at epoch $t$ if 
at $t=0$ it was in state $i$, $p_{ij}$ are the 
elements of the $p'$ matrix (\ref{p'})
and $c_i^{-1}$ is the characteristic waiting time 
in site $i$ and it represents the average life time for an 
avalanche of size $i$ in BS model.
Intuitively the average time of an avalanche is a function
of the average number of sites with fitness less than $\lambda$
which is increasing with the avalanche width;
at criticality we propose a behavior $c_i^{-1}\approx i^{\chi} $
 for $i >0$ and  $c_0^{-1}=\alpha$ with $\alpha$ a given constant. 

An avalanche is now defined as an off time interval from the origin,
 whose probability distribution is independent of $\alpha$
 \cite{Feller2}.
The avalanche distribution function can be expressed as
\begin{equation}\label{ctav}
p_{av}(t)=\sum_{n=1}^\infty p(n)p_n(t)
\end{equation}
where $p(n)$ is the probability of an $n$ steps excursion out of the
 origin and
it is the first return probability distribution for the random walk
 defined
in section II; $p_n(t)dt$ is the probability of the first return to
 the origin
in the interval $t,t+dt$ after $n$ steps.
Intuitively we may say that for the $\lambda < 2/3$ the exponential 
decay of $p(n)$ will prohibit 
the long time avalanches and the average off time will be finite
 \cite{Feller2},
 while at $\lambda =2/3$  there is a qualitative
change; even $p_n(t)$ is decaying exponentially  the scale invariance 
of $p(n)$ for large $n$ leads to critical behavior. 

We have performed numerical simulation for the continuous time
  random walk 
at criticality for four scaling law of the parameters $c_i,
\quad c_i=\lambda i^\chi$ with $\chi=1,\ 1.5,\ 2,\ 2.5$.
The  numerical values for the critical exponent $\tau$ (Table \ref{h})
decreases monotonically as $\chi$ increases.
This behavior is intuitively clear, the avalanches tends to last 
longer if the 
characteristics life times are growing faster and $\tau\rightarrow
 1$ if $ \chi\rightarrow\infty$.
  
\section{Conclusions}
We have proposed a new approach to Bak \& Sneppen  evolution model
based on the dynamics of the avalanche width, the critical 
exponents are equal to those found previously in the mean field 
solution \cite{Derrida}, in fact they are universal properties of the
one dimensional random walk, but the 
value  $\lambda_{critical}=2/3$ is very close to the numerical
 results reported in Ref. \cite{SB}, \cite{Grass}. 
Thereby we believe
that the dynamics of the avalanche width is carrying useful
 information on the critical behavior for this model.
The structure of the generating function (\ref{R}) is intimately
 connected with
critical behavior, the branch line appearing in $N \rightarrow 
\infty$ limit generates the algebraic decay 
for the probability distribution of the avalanches. 
A generating function  with a finite number of poles will lead, 
through formula (\ref{asym}),
to  an exponential decay of the avalanches probability distribution.
The continuous time random walk picture allows for  a more carefully 
analysis of the avalanche structure and it improves the $\tau$
 critical exponent
keeping $\lambda_{critical}$ at the same value; it also gives an
 intuitive decomposition of the the algebraic 
decay distribution of the avalanches 
in a convolution of Poisson distributed events.
 \section*{\center Acknowledgments}
The author thanks to  Amos Maritan for enlightening discussions
and for critical reading of the manuscript. 

\appendix\section*{}
We present the detailed calculation for the generating function of 
the avalanches
probability distribution. For this propose we shall use the special
 form of the matrix $\tilde p'$  obtained from 
the matrix $p'$ (\ref {p'}) removing the 
line and the row with index zero. This matrix has  equal diagonal
 elements
and we can write it as a linear combination of one-diagonal matrices
$I_i$ defined as follow:
\begin{equation}\label{Ii}
(I_i)_{kl}=\left\{
		\begin{array}{ll}
	        \delta_{k+i,l}&\ i\ge 0,\\
		\delta_{k,l+i}&\ i<0
	        \end{array}
\right .
\end{equation}
 $I_0$ being the identity matrix.
>From the definitions (\ref{Ii}) we can compote the commutator
$T^{(i)}=I_1I_{-i}-I_{-i}I_1\mbox{; for}\ i>0$ we have
\begin{equation}\label{com}
(T^{(i)})_{kl}=\delta_{i+1,1}
\end{equation}
and for $ij>0$ we have the  property
\begin{equation}\label{prod}
I_iI_j=I_{i+j}.
\end{equation}
The matrix  $\tilde p'$ can be expressed as:
\begin{eqnarray}
\tilde p'&=&\lambda ^2 I_1+2\lambda^2(1-\lambda)I_0+\sum_{i=1}
^{\infty}
(i+2)\lambda^2(1-\lambda)^{i+1}I_{-i}\nonumber\\
&=&\lambda^2I_1\sum_{i=0}^{\infty}(i+1)(1-\lambda)^iI_{-i}
=\lambda^2 I_1 A
\end{eqnarray}		   
where $A=\sum_{i=0}^{\infty}(i+1)(1-\lambda)^iI_{-i}$.
Using eq. (\ref{prod}) it is easy to compute the $n$-th power of this
matrix
\begin{equation}
A^n=\sum_{j=0}^{\infty}{2n+j-1\choose j}(1-\lambda)^jI_{-j}.
\end{equation}
>From eq. (\ref{com}) we can compute the commutator $T_n=
I_1A_n-A_nI_1$ which has only the first column non zero
\begin{equation}\label{Acom}
(T_n)_{kl}={2n+k+1\choose k}(1-\lambda)^k\delta_{l1}
\end{equation}
consequently,
\begin{equation}
(I_nT_n)_{j1}={3n+j-1\choose j+n}(1-\lambda)^{j+n}.
\end{equation}
All the previous mentioned properties lead us to the 
relation
\begin{equation}\label{I1A}
(I_1A)^n=I_nA^n-\sum_{i=0}^{n-2}I_iT_i(I_1A)^{n-i-2}.
\end{equation}
Eq. (\ref{I1A}) imply the following equation for the generating
matrix  $G(z)=\sum_{i=0}^{\infty}(\lambda^2I_1A)^iz^i$
\begin{equation}\label{Geq}
G(z)=F(z)-\sum_{i=1}^{\infty}I_iT_i\lambda^{2(i+1)}z^{i+1}G(z)
\end{equation}
where $ F(z)=\sum_{i=1}^{\infty}
I_iA^iz^i, \ z\ \mbox{complex number}$.
The sums  which are appearing in eq.(\ref{Geq}) can be done in the 
following way:
\begin{eqnarray}\label{u}
u_j(z)&=&\sum_{k=1}^{\infty}(I_kT_k)_{j1}\lambda^{2(k+1)}z^{k+1}
=\sum_{k=1}^{\infty}{3k+j-1\choose j+k}(1-\lambda)^{j+k}\lambda^{2(k+1)}
z^{k+1}\nonumber\\
&=&(1-\lambda)^{j-1}\xi^3\sum_{k=1}^{\infty}{1\over (k+j)!}(3k+j-1)
\dots 2k \xi^{2k -1}\nonumber\\
&=&(1-\lambda)^{j-1}\xi^3\sum_{k=1}^{\infty}\frac{1}{(k+j)!}
\frac{d^{k+j}}{d\xi^{k+j}}\xi^{3k+j-1}=(1-\lambda)^{j-1}\xi^3
\sum_{k=1}^{\infty}\frac{1}{2\pi i}
\oint_{\Gamma}d\eta \frac{\eta^{3k+j-1}} {(\eta-\xi)^{j+k+1}}
\nonumber\\
&=&(1-\lambda)^{j-1}\xi^3{1\over 2\pi  i}\oint_{\Gamma}d\eta 
{\eta^{j+2}
\over (\eta-\xi)^{j+1} (-\eta^3+\eta-\xi)}
\end{eqnarray}
where $\xi^2=(1-\lambda)\lambda^2z$.
We can perform the summation if $|\eta^3/(\eta-\xi)|<1$, this set 
is not empty
for $0<\xi <2/3\sqrt {1/3}$. There is a annulus with inner radius
 and external radius
obtained from the the positive solutions of the equation
 $(r+\xi)^3-r=0$;  
 more than that, one of the roots of the polynomial
$-\eta^3+\eta-\xi$ is inside of the minimal integration contour
 $\Gamma$ for $0<\xi <2/3\sqrt {1/3}$
and the other two are outside of the maximal integration contour.
Using the above mentioned properties of the matrices $\{I_k\}$ 
 (\ref{prod}) we can compute the elements of the 
matrix $I_nA^n$ which appear
in the expression of the generating matrix $F(z)$ : 
\begin{eqnarray}
(I_nA^n)_{1j}&=&{3n-j\choose n-j+1}(1-\lambda)^{n-j+1},\ \ \ 
(I_nA^n)_{21}={3n\choose n+1}(1-\lambda)^{n+1},\nonumber\\
(I_nA^n)_{2j}&=&(I_nA^n)_{1\, j-1}\quad j>1.\nonumber
\end{eqnarray}
Eq.(\ref{avp}) shows that we need to compute only the first 
two rows in the generating matrices $G(z)$ and $F(z)$.
The general formula for these matrix elements of $F(z)$ is
\[
F_{1j}(z)=\delta_{j1}+\sum_{n=j-1}^{\infty}(1-\lambda)^{n-j+1}
{3n-j\choose n-j+1}\lambda^{2n}z^n.
\]
The previous series can be summed following the same computational 
path as in eq.(\ref{u}).  If $j=1$ we get
 $$F_{1,1}(\xi)=1+{\xi\over 2\pi i}\oint_\Gamma 
{\eta^2\over (\eta-\xi)(-\eta^3+\eta-\xi)},$$
for $j>1$ we have the following expression:
\[
F_{1j}(\xi)={\xi\over (1-\lambda)^{j-1}}{1\over 2\pi i}
\int_\Gamma {\eta^{2j}\over -\eta^3+\eta-\xi},
\]
 $F_{2j}(z)=F_{1j-1}(z),\ j>1$, because $I_nA^n$ has equal 
elements on diagonals, and by direct calculation 
\[
F_{21}(z)=\sum_{n=1}^{\infty}{3n\choose n+1}(1-\lambda)^{n+1}z^n
={1-\lambda\over 2\pi i}\xi\oint_{\Gamma}{\eta^3\over (\eta-\xi)^2
(-\eta^3+\eta-\xi)}.
\]
In all the above formula the contour $\Gamma$ is the same as the 
that one used in eq.(\ref{u}) and $\xi^2=(1-\lambda)\lambda^2z$.
Solving eq.(\ref{Geq}) we obtain for the first two rows the 
solutions in
terms of previously computed functions $u_1(z),u_2(z),F_{1j}(z),
F_{2j}$: 
\begin{eqnarray}
G_{1,j}(z)&=&{F_{1j}(z)\over 1+u_1(z)},\nonumber\\
G_{2,j}(z)&=&F_{2j}(z)-{u_2(z)\over 1+u_1(z)}F_{1j}(z).\nonumber
\end{eqnarray}
Residue theorem allow us to compute the generating functions in term
 of the
third solutions of the polynomial $-\eta^3+\eta-\xi$, $\eta_3(\xi)$, 
that solution which
lies inside of the integration contour $\Gamma$ in the above 
integrals.
 \begin{eqnarray}\label{G}
G_{1j}(\xi)&=&{1\over (1-\lambda)^{j-1}}{\eta_3(\xi)^{2j}\over\xi^2}
\qquad i\ge 1,\\
G_{2j}(\xi)&=&-{1\over (1-\lambda)^{j-2}}\Bigl({1\over \xi^2}-2\Bigr)
{\eta_3(\xi)^{2j}\over
\xi^2}\qquad j\ge 1;
\end{eqnarray}
with $\xi^2=\lambda^2(1-\lambda)z$ and
\begin{equation}\label{eta}
\eta_3(\xi)=-{1-i\sqrt 3\over 2^{2/3}(-27\xi+\sqrt{729\xi^2-108})
^{1/3}} 
-{(1+i\sqrt 3)(-27\xi+\sqrt{729\xi^2-108})^{1/3}\over 6\> 2^{1/3}}.
\end{equation}
>From eq. (\ref{avp}) for the avalanches probability distribution one 
can write the generating function:
\begin{eqnarray}
R(z)&=&(1-\lambda)^2z+z^2p_{01}\sum_{i=1}^{\infty}G_{1i}(z)p_{i0}+
z^2p_{02}\sum_{i=1}^{\infty}G_{2i}(z)p_{i0}\nonumber\\
&=&(1-\lambda)^2z+2\lambda(1-\lambda)z^2\sum_{i=1}^{\infty}
{1\over 1+u_1(z)}F_{1i}(z)
(1-\lambda)^{i+1}(1+(i+1)\lambda)\nonumber\\
&&+z^2\lambda^2\sum_{i=1}^{\infty}(F_{2i}(z)-{u_2(z)\over 1+u_1(z)}
F_{1i}(z))
(1-\lambda)^{i+1}(1+(i+1)\lambda).\nonumber\\
\end{eqnarray}
The series which are appearing above can be summed and the closed
 expression
for generating function reads:
\begin{eqnarray}\label{R}
R(\xi(z))=-2{1-\lambda\over \lambda}\xi^2+{2(1-\lambda)\over \lambda^3}
\xi^2\eta_3(\xi)^2\biggr (1+2\lambda +{\eta_3(\xi)^2\over
 1-\eta_3(\xi)^2}
\Bigr (1+\lambda{3-2\eta_3(\xi)^2\over 1-\eta_3(\xi)^2}\Bigl )\biggl )
\nonumber\\
+{1-\lambda\over\lambda^2}(1-2\xi^2)\eta_3(\xi)^2\biggr (1+2\lambda+
(1+3\lambda)\eta_3(\xi)^2+{\eta_3(\xi)^4\over 1-\eta_3(\xi)^2}
\Bigr (1+
\lambda {4-3\eta_3(\xi)^2\over 1-\eta_3(\xi)^2}\Bigl )\biggl ).
\end{eqnarray}
The study of asymptotic behavior of the avalanches probability
 distribution is easily obtained by
studying the behavior of the derivative of 
$G_{1j}(z)$ and $G_{2j}(z)$ functions.
The summation appearing in $R(z)$ changes only the amplitude
 for the leading term.
It is crucial for the algebraic behavior  the existence of 
the branch line in the complex 
plane for the generating functions $G_{ij}(z)$  which we choose 
on the real axis from
 $(108/729)\lambda^2(1-\lambda)$ to $\infty$.

\begin{table}
\setdec 0.00
\caption{The numerical values of the critical exponent $\tau$ for four
values of $\chi$}
\begin{tabular}{cc}
$\chi$ & $\tau$ \\
\tableline
$1$   & $1.27\pm 0.01$ \\
$1.5$ & $1.19\pm 0.01$ \\
$2$   & $1.14\pm 0.01$ \\
$2.5$ & $1.01\pm 0.01$\\ 
\end{tabular}
\label{h}
\end{table}
\end{document}